\documentclass[12pt,eqsecnum,preprint]{aastex}
\title{X-ray Diagnostics of Broad Absorption Line Quasar Geometry}
\author{Brian Punsly}
\affil{4014 Emerald Street No.116, Torrance CA, USA 90503 and
International Center for Relativistic Astrophysics,
I.C.R.A.,University of Rome La Sapienza, I-00185 Roma, Italy}
\email{brian.m.punsly@boeing.com or brian.punsly@gte.net}\and
\author{Sebastian Lipari}
\affil{Cordoba Observatory and CONICET, Laprida 854, 5000 Cordoba,
Argentina} \email{lipari@mail.oac.uncor.edu}
\begin{document}
\begin{abstract}A new generation of sensitive X-ray measurements are
indicating that the existence of X-ray attenuation column
densities, $N_{H}>10^{24}\mathrm{cm}^{-2}$ is quite common amongst
broad absorption line quasars (BALQSOs). This is significant to
the geometry of the broad absorption line (BAL) outflow. In
particular, such an X-ray shield also shields equatorial accretion
disk winds from the UV, thereby preventing high velocity
equatorial outflows from being launched. By contrast, bipolar
winds initiated by continuum radiation pressure from the funnel of
a slim accretion disk flare outward (like a trumpet) and offer
vastly different absorbing columns to the X-ray and UV emission
which are emitted from distinct regions of the disk, $\sim 6M$ and
$\sim 10M-40M$, respectively (where $M$ is the radius of the black
hole). Recent numerical work indicates that it is also possible to
launch bipolar outflows from the inner regions of a thin disk. The
recent discovery with VLBI that the Galactic analog of a BALQSO,
the X-ray binary Circinus X-1 (with high velocity P Cygni X-ray
absorption lines) is viewed virtually along the radio jet axis
(and therefore along the spin axis of the black hole and the
normal to the accretion disk) has rekindled interest in the
bipolar models of BALQSOs. We explore this possibility by studying
the nearest BAL QSO, MRK 231. High resolution 2-D optical
spectroscopy and VLBI mappings of the radio jet axis indicates
that the BAL outflow is parallel to the parsec scale radio jet.
\end{abstract}
\section{Introduction} It is of great physical significance that about
10\% - 15\% of quasars show broad absorption line (loosely defined
as absorbing gas that is blue shifted at least 5,000 km/s relative
to the QSO rest frame and displaying a spread in velocity of at
least 2,000 km/s) outflows \citep{wey97}. It is widely believed
that all radio quiet quasars have BAL flows, but the designation
of a quasar as a BALQSO depends on whether the line of sight
intersects the solid angle subtended by the outflow. The standard
model of quasars is one of a hot accretion flow onto a black hole
and a surrounding torus of molecular gas \citep{ant93}. A very
popular model of the BALQSOs is an equatorial wind driven from the
luminous disk that is viewed at low latitudes, just above the
molecular gas \citep{mur95}. Alternatively, one can have a bipolar
BAL flow that conforms with the observations \citep{pun00,pro04}.
However, BALQSOs are so distant that direct imaging of the BAL
region is beyond the resolution of current optical telescopes.
Thus, much of the discussion of BAL geometry is based on deductive
reasoning. In this Letter, we implement observations of nearby BAL
systems in order to achieve the high resolution necessary to
determine the geometry directly. In particular, we discuss new
observations of the nearest BALQSO MRK 231.
\par Recently, new studies of the X-ray binary Circinus X-1 has
changed our perspective. Circinus X-1 has P Cygni absorption
profiles in the X-rays associated with a high velocity broad
absorption line flow. Thus, the accretion disk of Circinus X-1 has
been considered a Galactic example of a broad absorption line
quasar \citep{bra01}. Radio observations of Circinus X-1 have now
indicated jet outflow velocities of 15 c \citep{fen04}. This can
only occur for a relativistic flow that is viewed within a few
degrees of the axis of the jet \citep{lin85}. The Galactic example
of a broad absorption line quasar is therefore viewed pole-on.
\par The findings on Circinus X-1 raise the question, are
BAL QSOs being viewed close to the black hole spin axis? We begin
this empirical analysis of BAL geometry by studying the wealth of
new X-ray data on BALQSOs. A picture seems to be emerging that
shows Thompson thick absorbing columns to the X-ray source. In the
most realistic models of equatorial outflows, \citet{mur95,pro00},
Thompson thick columns are catastrophic because they block the UV
photons form accelerating the BAL region outward. Thus, the new
X-ray observations indicate that high velocity equatorial BAL
winds as we know them can not be attained in many BALQSOs. In
order to determine the geometry of a true BALQSO, one must turn to
the nearest example in order to achieve adequate resolution, MRK
231. Fortunately, MRK 231 also has a radio jet that is powerful
enough to be imaged on pc scales. In this Letter, we combine VLBI
mapping of the jet with 2-D spectroscopy of MRK 231 to show that
both the radio jet and the BAL region flow outward from the
nucleus along PA $\approx -120^{\circ}$.
\section{X-ray observations of BALQSOs} The X-ray flux from the
quasar tends to over-ionize gas the atoms in a putative BAL
outflow, preventing resonant absorption and the resultant line
driving force. In a breakthrough paper of \citet{mur95}, it was
shown that if the quasar X-rays were shielded, a line driven wind
was possible from the accretion disk which they argued was
preferable to an outflow of dense BAL clouds and the associated
confinement problems. The earliest X-ray observations of BALQSOs
supported the notion of X-ray shielding gas. ROSAT soft X-ray
measurements indicated that BALQSOs were X-ray suppressed relative
to other quasars \citep{gre96}. Numerically, further evidence was
found in \citet{pro00} in which models naturally produced a shield
of X-ray emitting gas between the central radiation source and the
high velocity winds. In this section, we consider the most recent
X-ray observations and their application to the details of the
shielding gas.
\subsection{Observing X-ray Absorption Columns}
The ROSAT observations of BALQSOs indicated that X-ray absorbing
columns $N_{H}>10^{22}\mathrm{cm}^{-2}$ were quite typical
\citep{gre96}. However, as observations at higher energy, higher
sensitivity and higher resolution have become available estimates
of these absorbing columns have increased dramatically. This
essentially is a consequence of the fact that at
$N_{H}=10^{23}\mathrm{cm}^{-2}$ the absorbing column is
essentially black to soft X-rays \citep{mur95}. One needs higher
energy X-ray measurements to learn more. Secondly, the fluxes from
the BALQSO nucleus are so attenuated that they can be swamped by
any nearby X-ray source. Thus, high resolution telescopes such as
XMM are valuable. We illustrate these points with a couple of
examples.
\par The prototypical BALQSO is PHL5200. Deep ASCA observations
indicated an absorbing column of $N_{H}\sim 5\times
10^{23}\mathrm{cm}^{-2}$ \cite{mat01}. However, deeper high
resolution imaging with XMM in \cite{bri02} indicated that the
poor resolution of ASCA allowed X-rays from a background radio
source to flood the detectors. Only a small percentage of the
total X-ray flux were attributable to PHL 5200. There were not
even enough photons detected in $\sim 46$ ksec to determine the
spectral properties and the absorption. One thing that is certain,
the absorbing column is clearly significantly larger than the ASCA
value of $N_{H}\sim 5\times 10^{23}\mathrm{cm}^{-2}$ and therefore
Thompson thick.
\par Next we review the X-ray data on MRK 231, the nearest BALQSO.
Using ASCA and ROSAT observations, \citet{tur99}, determined an
absorbing column of $N_{H}\sim 2\times 10^{22}\mathrm{cm}^{-2}$ to
$N_{H}\sim 10^{23}\mathrm{cm}^{-2}$ depending on whether the
absorbing material is neutral or ionized. The ASCA data were
analyzed in \citet{mal00} who deduced that the direct observation
of the nuclear X-rays was blocked by an optically thick absorber
and we observe a strong reflected and scattered component through
a neutral column density of $N_{H}\sim 3\times
10^{22}\mathrm{cm}^{-2}$. Similarly, from Chandra data
\citet{gal02} find a best fit to the data if the central source is
blocked by a Compton thick absorber, $N_{H}\sim
10^{24}\mathrm{cm}^{-2}$ and we detect scattered X-rays through
multiple lines of sight through absorbing columns of $N_{H}\sim
10^{21}\mathrm{cm}^{-2}$ - $N_{H}\sim 6\times
10^{23}\mathrm{cm}^{-2}$. Finally, hard X-ray BeppoSAX
observations of MRK 231 combined with XMM deep observations have
been made by \citet{bra04}. The high energy coverage combined with
the 20 ks XMM observation provides the best determination of
spectral parameters to date. They find that $N_{H}\sim 2\times
10^{24}\mathrm{cm}^{-2}$. Thus, as the measurements have improved,
the column density estimates have increased.
\par We also note the deep ASCA observation of PG0946+301 in
\citet{mat00}. After previous non-detections only yielded upper
limits to $N_{H}$, this deep observation was best fit with a
Thompson thick column of $N_{H}\gtrsim 10^{24}\mathrm{cm}^{-2}$.
The large column densities that have been deduced from the X-ray
data are clearly only lower limits. Once the absorbing column is
black, small amounts of scattered X-rays or X-rays leaking through
a patchy absorber (small domains with
$N_{H}\sim10^{22}\mathrm{cm}^{-2}$ to $N_{H}\sim
10^{23}\mathrm{cm}^{-2}$) will dominate the X-ray flux and $N_{H}$
of the bulk of the absorbing column will be greatly
underestimated. Therefore, in view of the recent X-ray data it is
now apparent that many BALQSOs are likely to have X-ray absorbing
columns with Compton optical depths considerably larger than one.
\subsection{The Implications for Disk Wind Models}In order
for an equatorial flow to be launched from an accretion disk, the
wind must emanate at large distances from the hole, $r\geq
10^{16}$ cm, so that the radiation source appears as a flux of
photons roughly parallel to the accretion disk surface. This
allows momentum transfer along the equatorial direction, driving
the equatorial BAL wind. As the wind is launched closer and closer
to the source it looks more and more bipolar, since the photon
flux attains a large vertical component (see \citet{pro00} and
references therein).
\par A second necessary ingredient for the geometry is a region of
X-ray shielding gas in order to insure the proper ionization state
is achieved at $r\geq 10^{16}$ cm so that lithium like atoms are
available for the line driving force to work efficiently
\citep{mur95}. The discovery of the large X-ray absorbing columns
in BALQSOs has been heralded as the likely verification of the
shielding gas at the inner edge of the wind and a triumph of the
model \cite{gal02}. However, this is not really shielding gas, but
filtering gas: it must filter out the X-rays and allow the UV
emission to pass through with small attenuation. By geometric
considerations, the UV and X-ray emitting regions are well inside
of the distant wind zone and the shield \citep{pro00}. Thus, if
the X-rays are shielded, so are the UV photons. For this reason
\citet{pro00} states explicitly that for $N_{H}\gtrsim
10^{24}\mathrm{cm}^{-2}$, "the line force form the central engine
could be so much reduced that it may not be able to accelerate the
gas to high radial velocities and no strong wind will be
produced." Thus, the large column densities that have been found
are problematic for equatorial models of BALQSO winds. However,
there are disk wind models of bipolar outflows that do not suffer
these same problems. Bipolar outflows are ejected from the funnels
of slim disks in \citet{pun00} and the work of \citet{pro04}
strongly suggest that bipolar winds can be driven from a thin
accretion disk at the relatively small radii.
\section{The BAL Outflow Geometry of MRK 231}Bipolar models of
BALQSOs do not suffer the UV attenuation problem that arises from
the large X-ray absorbing columns \citep{pun99,pun00}. The wind is
launched by continuum radiation pressure and subsequently boosted
by line driving forces farther out. Thus, near its base the
outflow is a hypersonic under-expanded jet that flares outward
like a trumpet. Thus, regions at different cylindrical radii on
the surface of the disk (i.e., such as the X-ray region, $r\sim 6M
- 7M$, and the UV region, $r\sim 10M - 40M$, where M is the radius
of a rapidly rotating central black hole) are exposed to different
absorbing columns of wind outflow. The X-rays typically pass
through $N_{H}\sim 10^{24}\mathrm{cm}^{-2}$ to $N_{H}\sim
10^{25}\mathrm{cm}^{-2}$ and the UV passes through $N_{H}\sim
10^{23}\mathrm{cm}^{-2}$ \citep{pun00}. Combining this with the
observation of a polar BAL outflow in Circinus X-1, it is of
interest to see if any BALQSO can be shown to have a polar BAL
wind. The best possibility for success is to choose the nearest
BALQSO in order to get adequate resolution, MRK 231.
\subsection{VLBI Observations of MRK 231}High resolution VLBI can be used to determine a radio jet axis at
1pc and give an estimate of the spin axis of the black hole and
the therefore the normal to the accretion disk. MRK 231 has been
observed with VLBI by \citet{tay94,ulv99,ulv00,lon03}. The highest
resolution measurements have been made at 15 GHz in
\citet{ulv99,ulv00} with a resolution of about 0.9 pc. A jet like
structure propagates at PA = $-112^{\circ}$ in one epoch and PA =
$-115^{\circ}$ in another epoch. The large variability of the
northeast component (a factor of 2.5 in 1.8 years) allows its
identification as the radio core. This identification of the core
is consistent with our 2-D spectroscopy.
\subsection{2-D Optical Spectroscopy of MRK 231} Optical
spectroscopy was performed on MRK 231 with the 4.2 m William
Herschel Telescope in \citet{lip05}. The detector was an array of
fiber optics and a spectrum was reduced from each of the fiber
optic responses. Each fiber optic captured light from a sector of
the sky 0.45" in radius. The result is a 2-D mosaic of spectra
that covers the central region of MRK 231, 10 kpc x 13 kpc in
extent.
    \par The results of the spectroscopy that are relevant are
    plotted in figure 1. We are interested in features representing an outflow
    and the blue wing of the broad H$\alpha$ line provides some clues. The upper spectrum is from the nuclear
    fiber optic and encompasses the region $r<366$ pc, $H_{0}=75$
    km/s/Mpc. The lower plot is from the adjacent fiber optic at PA
    = $-120^{\circ}$, $490 \mathrm{pc}<r<1220\mathrm{pc}$ from the nucleus. Note the
    broad emission line feature that is superimposed on the blue
    wing of the H$\alpha$ line. This feature exists in the nuclear
    fiber optic and is more pronounced in the adjacent fiber optic at
    PA = $-120^{\circ}$. There is no evidence of this blue bump in the
    spectra of any other of the fiber optics that are adjacent to
    the nucleus (the spectra at all of the fiber optic positions
    can be found in \citep{lip05}). The conclusion is that this is
high velocity
    H$\alpha$ emitting gas that is flowing at PA
    $\approx -120^{\circ}$. The gas is actually radiating more blue shifted
    photons from within $r<366$ pc, but this is swamped by the
    elevated background emission.
    \par One can estimate the peak velocity of the blue shifted
    H$\alpha$ by extrapolating the background broad line wing.
    This is indicated by the dashed lines in figure 1. The peaks
    of the emission relative to the background wing are indicated
    by daggers. The extrapolation of the continuum is somewhat
    arbitrary and we have chosen are best fit so that the two
    spectra have similar background blue wings. With this fit,
    we use the peak of H$\alpha$ at $\lambda=6839\AA$ from
    \citet{lip94} and the cosmological redshift of z=0.042 to find
    the relativistic outflow velocities associated with
    the blue shifted peaks of 4,729 km/s in the
    nucleus and 4,506 km/sec at PA = $-120^{\circ}$ \citep{lig75}. Note that there
    is some uncertainty in the estimate of the PA = $-120^{\circ}$
    peak due to the emergence of narrow line emission that
    contaminates the broad profile. We fit the background blue wings in a variety of
    ways that modify the velocity estimate associated with the peak of the nuclear
    outflow (varies within the range of 4,500 km/sec - 5,000 km/sec depending on the
    background fit) and for the PA = $-120^{\circ}$ spectra (varies from
    4,300 km/sec - 4,700 km/sec, depending on the background fit).
    \par As mentioned above the narrow line emission at PA =
    $-120^{\circ}$ adds some uncertainty to the estimate of the
    velocity of the blue bump emitting gas. It could be
    responsible for the slight disagreement with the nuclear blue
    bump velocity, since the narrow lines contribute flux to the red
    side of the blue bump. If one goes to the next fiber optic
    out along PA = $-120^{\circ}$ located between 1340 pc and 2030 pc
    from the nucleus, the narrow line emission becomes very strong
    \citep{lip05}. None of the other fiber optics show this strong emission.
    The generation of a strong narrow line region is expected from
    jet propagation in the ISM \citep{bic98}. In fact, a similar
    spectral effect has been attributed to bow shocks in the
    Circinus galaxy \citep{vei97}. Thus, it seems clear that the outflow, initially traced by
    the inner VLBI jet, extends $>1$ kpc at PA $\approx -120^{\circ}$. This direction
    validates the variable VLBA component as the nucleus (as
    opposed to PA $\approx 60^{\circ}$).
    \subsection{The Relationship to the BAL Gas} MRK 231 is
    notable for the existence of many strong low ionization
    broad absorption lines. It is particularly noteworthy that the
    primary Na I D doublet, He I $\lambda$3889 I, Ca II K $\lambda$3934 and
    Ca II H $\lambda$3968 broad absorption lines occur with velocities
    4,590 km/sec - 4,670 km/sec \citep{lip94,mar97}. There is H$\alpha$ emitting gas
    in the nuclear region that also flows at 4,690 km/sec. Either
    this is a coincidence or the two flows are one and the same.
    The latter seems more plausible to us and therefore, we
    conclude that the low ionization BAL gas is flowing outward at PA $\approx -120^{\circ}$, parallel to the projection of the jet axis on the
    sky plane to within $10^{\circ}$. Detailed studies of other low ionization
    BALQSO place the BAL region $\sim$ 200 pc -700 pc from the nucleus \citep{dek01,dek02}.
    However, since the blue bump is the most luminous in the nuclear fiber optic, the
    BAL flow can be located anywhere within the nuclear region,
    $r<$ 366 pc and need not be at such large distances.
    \section{Conclusion}In this Letter, we have discussed how the
    large observed X-ray absorbing columns in BALQSOs pose an apparent catastrophic problem
    for the currently realistic existing equatorial BAL wind models. We also
    point out that this is not an issue for bipolar flows. Evidence is
    presented from Circinus X-1 and MRK 231 that bipolar
    BAL flows seem to exist. There is likely many sources of BAL flows
    in quasars. This is particularly true for IR QSOs and high redshift QSO which are likely
    to be young objects \citep{lip05}. Our analysis indicates that bipolar BAL flows
    could be common.
    \par Motivated by our findings it has become apparent that efforts to create new
    numerical models that are capable of driving bipolar winds
    from accretion disks is essential to our understanding of BALQSOs. One promising model
    is the disk wind model of \citet{pro04}. At relatively small radii, it appears
    that bipolar winds can be launched. At present, it is difficult to compare the
    anticipated observed properties of these inner-disk driven winds with
    funnel driven outflows since the published model is merely an
    example of a broad class of models that was not optimized to
    produce for bipolar winds. The existing model has some
    apparent observable properties due to the choice of a disk from \citep{sha73}
    with the far UV emitting region very distant from the
black hole at $r<150M$ that creates a large extreme UV (EUV)
source ($\lambda=100\AA$ to $\lambda=1000\AA$) in the inner
regions of the disk. This source over-produces the flux in the EUV
by 2 orders of magnitude compared to composite HST quasar spectra
\citep{tel02,zhe97}. There is a large scatter in the quasar
spectral shapes used in these composites, so it is possible that
some individual spectral energy distributions could peak deep into
the EUV. But, this is clearly not typical. At the present level of
modeling, it is not clear if this large EUV excess is an expected
characteristic of bipolar winds from the inner disk. It might be
that the parameters of future models will not require such an EUV
excess to occur. It seems that a high priority in this field is to
reproduce these simulations with a different disk model, perhaps a
larger central black hole mass about a rapidly spinning black hole
that will remove the EUV excess and perhaps enhance the bipolar
wind production from the inner disk.
    \acknowledgments
    We would like to thank Kirk Korista, Timothy Kallman and
    Robert Antonucci for input that helped with the revision of
    this manuscript. The comments of an anonymous referee greatly improved the final presentation of this discussion.

\begin{figure}
\plotone{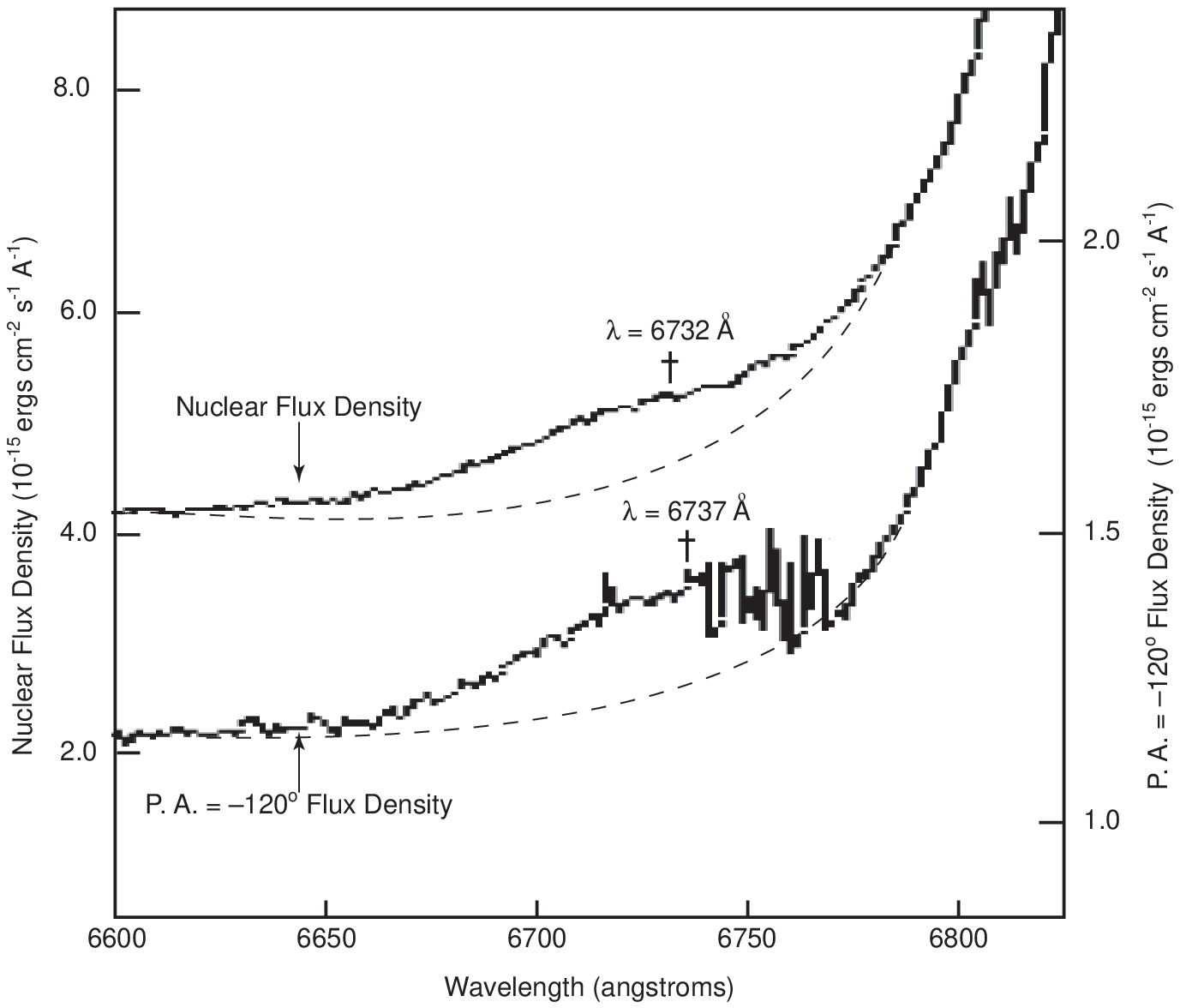}
    \caption{The 2- D spectra from the William Herschel Telescope with the INTEGRAL fiber system
     of the blue wing of the H$\alpha$ emission from the nuclear
    fiber optic (upper plot) and the adjacent fiber optic at PA =$-120\deg$ (lower plot).
    The left axis is the flux density of the nucleus and the right axis is the flux density
    from the fiber optic at PA = $-120^{\circ}$. The nuclear flux is the from a region,
     $r<366$ pc. The fiber optic at PA = $-120^{\circ}$ is a distance,
      $490\mathrm{pc}<r<1220\mathrm{pc}$ from the nucleus.}
    \end{figure}
\end{document}